\documentclass[12pt]{article}
\usepackage[english]{babel}
\usepackage[utf8]{inputenc}
\usepackage[english]{babel}

\usepackage{amsfonts}

\begin{document}
\baselineskip 21pt

\title{The spherically symmetric gravitational field}
\author{Sergey Gubanov\/\thanks{s.yu.gubanov@inbox.ru}}
\maketitle

\begin{abstract}
The general solution of the system of General Relativity equations has been found
for isotropic Universe with the flat spatial distribution and synchronized time
taking into account a perfect dust and the cosmological constant.
Schwarzschild, Friedmann and Einstein--de Sitter solutions
(as well as all of their fusion with each other)
are special cases of the found general solution.
A method of generating an infinite number
of Tolman's like solutions has been found.
Exact solutions has been found
for the spherically symmetric gravitational field of perfect dust clouds
in the expanding Universe filled with radiation.
A system of ordinary differential equations has been obtained
for the spherically symmetric gravitational field
of perfect dust clouds in the expanding Universe
filled with radiation and nonrelativistic gas.
A system of equations has been obtained
for the spherically symmetric gravitational field
of ultrarelativistic celestial body explosion (supernova, quasar).
The problem of a negative density of a perfect dust cloud in General Relativity
has been considered.
\end{abstract}

\noindent
{\bf Key words:\/} gravitation, cosmology, exact solutions, General Relativity.

\noindent
{\bf PACS codes:\/} 04.20.Jb, 04.70.Bw

%***************************************************************
\section{Introduction}
\noindent
Let us consider a model of the isotropic Universe
with flat spatial distribution
and global synchronized time.
In this case the metric is dependent only from one function
$V(t, r)$ and is given by the following formula
(see below the section ''Space-Time structure''):
\begin{equation}
\label{metric1}
ds^2 = dt^2 - \left( dr - V(t,r) \, dt \right)^2 
- r^2 \left( d\theta^2 + \sin^2 (\theta) \, d \varphi^2 \right).
\end{equation}
By appropriate coordinate transformation, 
it is possible to transform to the metric~(\ref{metric1}) 
following well known solutions: 
Schwarzschild solution~\cite{Schwarzschild1915};
Reissner--Nordström solution~\cite{Reissner1916}, \cite{Nordstrom1918};
Einstein--de Sitter solution~\cite{deSitter1917};
Friedmann--Lemaître--Robertson--Walker (FLRW) metric with
flat spatial distribution~\cite{Friedmann1922};
Tolman solution in the case of flat spatial distribution~\cite{Tolman1934}. 
For example,
the Schwarzschild--Reissner--Nordström--Einstein--de Sitter metric
\begin{equation}
ds^2 = \left( 1 -  \frac{2 \kappa M}{r} + \frac{\kappa Q^2}{r^2} - \frac{4}{9} \lambda^2 r^2 \right) d\tilde{t}^2 
- \frac{dr^2}{1 -  \frac{2 \kappa M}{r} + \frac{\kappa Q^2}{r^2} - \frac{4}{9} \lambda^2 r^2 }
- r^2 \left( d\theta^2 + \sin^2 (\theta) \, d \varphi^2 \right)
\end{equation}
can be transformed to the metric~(\ref{metric1})
\begin{equation}
ds^2 = dt^2 
- \left( dr \pm \sqrt{\frac{2 \kappa M}{r} - \frac{\kappa Q^2}{r^2} + \frac{4}{9} \lambda^2 r^2  } \, dt  \right)^2
- r^2 \left( d\theta^2 + \sin^2 (\theta) \, d \varphi^2 \right)
\end{equation}
by Painlevé--Gullstrand~\cite{Painleve1921}, \cite{Gullstrand1922} coordinate transformation:
\begin{equation}
d\tilde{t} = dt \pm \frac{ \sqrt{ \frac{2 \kappa M}{r} - \frac{\kappa Q^2}{r^2} + \frac{4}{9} \lambda^2 r^2 } }{1 - 
\frac{2 \kappa M}{r} + \frac{\kappa Q^2}{r^2} - \frac{4}{9} \lambda^2 r^2  } dr,
\qquad
\frac{2 \kappa M}{r} - \frac{\kappa Q^2}{r^2} + \frac{4}{9} \lambda^2 r^2 \ge 0.
\end{equation}
The FLRW metric with flat spatial distribution
\begin{equation}
\label{metricFLRW}
ds^2 = dt^2 
-  a^2(t)  \left(d\tilde{r}^2
+ \tilde{r}^2 d\theta^2 + \tilde{r}^2 \sin^2 (\theta) \, d \varphi^2 \right)
\end{equation}
can be transformed to the metric~(\ref{metric1})
\begin{equation}
\label{fridman}
ds^2 = dt^2 
- \left( dr - r \frac{\dot{a}(t)}{a(t)} \, dt  \right)^2
- r^2 \left( d\theta^2 + \sin^2 (\theta) \, d \varphi^2 \right)
\end{equation}
by radial $\tilde{r}$ coordinate transformation:
\begin{equation}
\label{radiusFLRW}
\tilde{r}(t, r) = \frac{r}{a(t)}.
\end{equation}
The Tolman metric has two arbitrary functions $M(\xi)$ and $W(\xi)$:
\begin{equation}
ds^2 = dt^2 - \left( \frac{ \frac{\partial r}{\partial \xi} d \xi }{W(\xi)} \right)^2
- r^2(t, \xi) \left( d \theta^2 + \sin^2(\theta) d \varphi^2 \right).
\end{equation}
There the function $r(t, \xi)$ is a solution of the following equation:
\begin{equation}
\frac{\partial r}{\partial t} = \pm 
\sqrt{ \frac{2 \kappa M(\xi)}{r} + \frac{4}{9}\lambda^2 r^2 + W^2(\xi) - 1 }.
\end{equation}
When we take the function $r(t, \xi)$ as radial coordinate
the metric take the following form
\begin{equation}
\label{tolman}
ds^2 = dt^2 -
\left( \frac{ dr \mp \sqrt{ \frac{2 \kappa M}{r} + \frac{4}{9}\lambda^2 r^2 + W^2 - 1 } \, dt 
}{ W } \right)^2
- r^2 \left( d \theta^2 + \sin^2(\theta) d \varphi^2 \right),
\end{equation}
and in the case $W=1$ we get the metric~(\ref{metric1}).
The case $W \ne 1$ considered further.

\noindent
Because well known metrics listed above are transformable to the metric~(\ref{metric1}),
it is interesting to find for the metric~(\ref{metric1}) the general solution of the system of equations of General Relativity
taking into account a perfect dust and the cosmological constant:
\begin{equation}
\label{eq1}
G_{\mu \nu} - \frac{4}{3}\lambda^2 g_{\mu \nu} = 8 \pi \kappa \, T_{\mu \nu}.
\end{equation}
The general solution $V(t, r)$ of this system of equations
have to give Schwarzschild, FLRW and Einstein--de Sitter
solutions in particular cases and all of their fusion with each other.

%*************************************************************
\section{Dust solutions}

\noindent
In this section the general solution of the system~(\ref{eq1})
for the metric~(\ref{metric1}) is obtained
taking into account a perfect dust and the cosmological constant.

\subsection{The Einstein tensor}
\noindent
Non-zero components of the Einstein tensor
for the metric~(\ref{metric1}) are listed below
\begin{equation}
G_{t t} = - \frac{V}{r^2} \left( V^3 - 2 r V' (1 - V^2) + V (2 r \dot{V} - 1) \right),
\end{equation}
\begin{equation}
G_{t r} = \frac{2 V}{r^2} \left( r \dot{V} + \frac{1}{2} \left( r V^2 \right)' \right),
\end{equation}
\begin{equation}
G_{r r} = - \frac{2}{r^2} \left( r \dot{V} + \frac{1}{2}\left( r V^2 \right)' \right),
\end{equation}
\begin{equation}
G_{\theta \theta} = - r \left( r \dot{V} + \frac{1}{2} \left( r V^2 \right)' \right)',
\end{equation}
\begin{equation}
G_{\varphi \varphi} = - r \sin^2(\theta) \left( r \dot{V} + \frac{1}{2} \left( r V^2 \right)' \right)'.
\end{equation}
The point and the prime mark denotes differentiation with respect to $t$ and $r$, respectively.
The Einstein tensor definition adopted in~\cite{LL2} has been used.

%*************************************************************
\subsection{The stress--energy tensor of a perfect dust}
\noindent
The stress--energy tensor of a perfect dust is given by~\cite{LL2}:
\begin{equation}
\label{Trho}
T_{\mu \nu} = \rho \, u_{\mu} u_{\nu}.
\end{equation}
The four-velocity $u^{\mu}$ of the perfect dust must satisfy the following equations
\begin{equation}
\label{dusteq}
g_{\mu \nu} u^{\mu} u^{\nu} = 1, \quad 
u^{\mu} \left( \nabla_{\mu} u^{\nu} \right) = 0.
\end{equation}
The interesting for us solution of the system~(\ref{dusteq}) for the metric~(\ref{metric1}) is given by
\begin{equation}
\label{dustu}
u^{\mu} \frac{\partial}{\partial x^{\mu}} =
\frac{\partial}{\partial t}
+ V \frac{\partial}{\partial r}, \quad
u_{\mu} dx^{\mu} = dt.
\end{equation}
Taking into account~(\ref{dustu}) we obtain for the stress--energy tensor
\begin{equation}
\label{dustt}
T_{\mu \nu} \, dx^{\mu} dx^{\nu} = \rho \, dt^2.
\end{equation}
The continuity equation for the scalar mass density $\rho(t, r)$ is given by
\begin{equation}
\label{rhoEq}
\nabla_{\mu} T^{\mu \nu} = 0
\quad \to \quad
\nabla_{\mu} \left( \rho u^{\mu} \right) = 0
\quad \to \quad
\frac{\partial \rho }{\partial t}
+ \frac{1}{r^2} \frac{\partial}{\partial r} \left( r^2 \rho \, V \right) = 0.
\end{equation}
Note that a mass density $\rho(t, r) \ge 0$ is a scalar function,
an energy density $T_{t t} \ge 0$ is a tensor function,
they are equal $T_{t t} = \rho$ due to the solution~(\ref{dustu}) for the metric~(\ref{metric1}).

%*************************************************************
\subsection{The system of equations}
\noindent
Due to Hilbert identities~\cite{Hilbert1915}
the continuity equation~(\ref{rhoEq}) already has been contained in the system~(\ref{eq1})
with the right side~(\ref{dustt}).
Therefore the solution of the system~(\ref{eq1}) we can find as described below.
At the beginning we will find solution of the $t r$, $r r$, $\theta \theta$ and $\varphi \varphi$-equations
(they have zero right-hand side).
It is easy to see these equations are reduced to a single equation:
\begin{equation}
\label{eqV}
\frac{\partial V}{\partial t} 
+ \frac{1}{2 r} \frac{\partial}{\partial r} \left( r V^2 - \frac{4}{9} \lambda^2 r^3 \right) = 0.
\end{equation}
Due to~(\ref{eqV}) we have for the left-hand side of $t t$-equation
\begin{equation}
G_{t t} - \frac{4}{3}\lambda^2 g_{t t}
\equiv - \frac{2}{r} \frac{\partial V}{\partial t}.
\end{equation}
The $t t$-equation of General Relativity
\begin{equation}
\label{ttEq}
G_{t t} - \frac{4}{3}\lambda^2 g_{t t} = 8 \pi \kappa \rho
\end{equation}
now can be used as a definition for the density $\rho$ of the perfect dust:
\begin{equation}
\rho = - \frac{1}{4 \pi \kappa \, r} \frac{\partial V}{\partial t}.
\end{equation}
Because $\rho \ge 0$, only $\frac{\partial V}{\partial t} \le 0$ solutions allowed in General Relativity.
Since the equation~(\ref{eqV}) has no limitation for the sign of $\frac{\partial V}{\partial t}$ at all,
it is a {\it negative density problem} in General Relativity,
it has been considered below in the section ''Gravitational field equations''.

\subsection{General solution}
\noindent
The general solution $V_{[F]}(t, r)$ of the equation~(\ref{eqV}) is given by:
\begin{equation}
\label{generalSolution}
F\left( \sqrt{r} 
\left( \cosh(\lambda t) V_{[F]} - \frac{2\lambda r}{3} \sinh(\lambda t) \right),
\sqrt{r} \left( 
\sinh(\lambda t) V_{[F]} - \frac{2\lambda r}{3} \cosh(\lambda t) \right)
\right) = 0.
\end{equation}
In the case $\lambda = 0$ we obtain:
\begin{equation}
\tilde{F}\left( \sqrt{r} \, V_{[\tilde{F}]} ,
\sqrt{r} \left( t \, V_{[\tilde{F}]} - \frac{2 r}{3} \right) \right) = 0.
\end{equation}
Here $F(\alpha, \beta)$ is a differentiable arbitrary function of two arguments.

%*************************************************************
\subsection{A few examples of particular solutions}
\paragraph{Solution 1}
In the case $F(\alpha, \beta) = \alpha \pm \sqrt{2 \kappa M}$ we have:
\begin{equation}
\label{solution1}
V^{\pm} = \frac{2}{3} \lambda r \tanh(\lambda t) \pm \frac{1}{\cosh(\lambda t)} \sqrt{\frac{2 \kappa M}{r}}.
\end{equation}
\begin{equation}
\label{solution1Gtt}
G^{\pm}_{t t} - \frac{4}{3}\lambda^2 g^{\pm}_{tt} = \frac{2\lambda}{3 r \cosh^2(\lambda t)}
\left( - 2 \lambda r \pm 3 \sinh(\lambda t) \sqrt{\frac{2 \kappa M}{r}} \right).
\end{equation}
In the case $\lambda = 0$ we obtain:
\begin{equation}
\label{blackwhitehole}
V = \pm \sqrt{\frac{2 \kappa M}{r}}.
\end{equation}
\begin{equation}
G_{t t} = 0.
\end{equation}
It is the Schwarzschild solution in the Painlevé--Gullstrand coordinates.
Negative sign in~(\ref{blackwhitehole}) is corresponding to a black hole
and positive sign in~(\ref{blackwhitehole}) is corresponding to a white hole.
Due to the $t t$-equation of General Relativity~(\ref{ttEq}) and the expression~(\ref{solution1Gtt})
the density $\rho(t, r)$ is not a positive definite,
so the solution~(\ref{solution1}) does not always (everywhere) belong to General Relativity.

\paragraph{Solution 2}
In the case $F(\alpha, \beta) = \beta$ we have:
\begin{equation}
V = \frac{2}{3} \lambda r \coth(\lambda t),
\end{equation}
\begin{equation}
G_{t t} - \frac{4}{3}\lambda^2 g_{tt} = \frac{4 \lambda^2}{3 \sinh^2(\lambda t)}.
\end{equation}
In the case  $\lambda = 0$ we obtain:
\begin{equation}
V = \frac{2 r}{3 t},
\end{equation}
\begin{equation}
G_{t t} = \frac{4 }{3 t^2}.
\end{equation}
It is the Friedmann solution with flat spatial distribution.

\paragraph{Solution 3}
In the case $F(\alpha, \beta) = \alpha + \beta$ we have:
\begin{equation}
V = \frac{2}{3} r \lambda,
\end{equation}
\begin{equation}
G_{t t} - \frac{4}{3}\lambda^2 g_{tt} = 0.
\end{equation}
It is the Einstein--de Sitter solution.

\paragraph{Solution 4}
In the case $F(\alpha, \beta) = \alpha^2 - \beta^2 - 2 \kappa M$ we have:
\begin{equation}
\label{blackwhitehole2}
V^{\pm} = \pm \sqrt{\frac{2 \kappa M}{r} + \frac{4}{9}\lambda^2 r^2 },
\end{equation}
\begin{equation}
G^{\pm}_{t t} - \frac{4}{3}\lambda^2 g^{\pm}_{tt} = 0.
\end{equation}
It is the minimal fusion of the Schwarzschild solution and the Einstein--de Sitter solution.
Negative sign in~(\ref{blackwhitehole2}) is corresponding to a black hole
and positive sign in~(\ref{blackwhitehole2}) is corresponding to a white hole.

\paragraph{Solution 5}
In the case $F(\alpha, \beta) = \beta \mp \lambda R^{3/2}$ we have:
\begin{equation}
\label{solution5}
V^{\pm} = \frac{2}{3} \lambda r \coth(\lambda t)  \pm \frac{\lambda R}{\sinh(\lambda t)} \sqrt{\frac{R}{r}},
\end{equation}
\begin{equation}
\label{solution5Gtt}
G^{\pm}_{t t} - \frac{4}{3}\lambda^2 g^{\pm}_{tt} = \frac{4 \lambda^2}{3  \sinh^2(\lambda t)}
\left( 1 \pm \frac{3}{2} \cosh(\lambda t) \sqrt{\frac{R^3}{r^3}} \right).
\end{equation}
In the case $\lambda = 0$ we obtain the Burlankov solution~\cite{Burlankov2011}:
\begin{equation}
\label{VB}
V^{\pm} = \frac{2 r}{3 t} \pm \frac{R}{t} \sqrt{\frac{R}{r}},
\end{equation}
\begin{equation}
G^{\pm}_{t t} = \frac{4 }{3  t^2}
\left( 1 \pm \frac{3}{2} \sqrt{\frac{R^3}{r^3}} \right).
\end{equation}
It is the minimal fusion of the Schwarzschild solution
and the Friedmann solution with flat spatial distribution.
Negative sign in~(\ref{VB}) is corresponding to a black hole
and positive sign in~(\ref{VB}) is corresponding to a white hole in the Friedmann expanding Universe.
Due to the $t t$-equation of General Relativity~(\ref{ttEq}) and the expression~(\ref{solution5Gtt})
the density $\rho(t, r)$ is not a positive definite,
so the solution~(\ref{solution5}) does not always (everywhere) belong to General Relativity.

\paragraph{Solution 6}
In the case $F(\alpha, \beta) = A \alpha + B \beta$ we have:
\begin{equation}
\label{solution6}
V = \frac{2 \lambda r}{3} \, \frac{A \sinh(\lambda t) + B \cosh(\lambda t)}{A \cosh(\lambda t) + B \sinh(\lambda t)},
\end{equation}
\begin{equation}
\label{solution6Gtt}
G_{t t} - \frac{4}{3}\lambda^2 g_{tt} = \frac{4\lambda^2 \left( B^2 - A^2 \right)}{
3 \left( A \cosh(\lambda t) + B \sinh(\lambda t) \right)^2}.
\end{equation}
It is the minimal fusion of the Einstein--de Sitter solution
and the Friedmann solution with flat spatial distribution.
Note that cosmological singularity does not exist if $B=0$.
Due to the $t t$-equation of General Relativity~(\ref{ttEq}) and the expression~(\ref{solution6Gtt})
the density $\rho(t, r)$ is not a positive definite,
so the solution~(\ref{solution6}) does not always (everywhere) belong to General Relativity.

\paragraph{Solution 7}
In the case $F(\alpha, \beta) = A \alpha + B \beta \mp \sqrt{2 \kappa M}$ we have:
\begin{equation}
\label{solution7}
V^{\pm} = \frac{1}{A \cosh(\lambda t) + B \sinh(\lambda t)} \left( \pm \sqrt{\frac{2 \kappa M}{r}} +
\frac{2}{3} \lambda r \left( A \sinh(\lambda t) + B \cosh(\lambda t) \right) \right),
\end{equation}
\begin{equation}
\label{solution7Gtt}
G^{\pm}_{t t} - \frac{4}{3}\lambda^2 g^{\pm}_{t t} = 
\frac{4 \lambda^2 \left( (B^2 - A^2) \pm \frac{3}{2 \lambda r} \sqrt{\frac{2 \kappa M}{r}}
\left( A \sinh(\lambda t) + B \cosh(\lambda t) \right)
  \right) }{3 (A \cosh(\lambda t) + B \sinh(\lambda t))^2 }.
\end{equation}
It is the minimal fusion of the Schwarzschild solution, the Einstein--de Sitter solution
and the Friedmann solution with flat spatial distribution.
Due to the $t t$-equation of General Relativity~(\ref{ttEq}) and the expression~(\ref{solution7Gtt})
the density $\rho(t, r)$ is not a positive definite,
so the solution~(\ref{solution7}) does not always (everywhere) belong to General Relativity.

\paragraph{Solution 8}
In the case $F(\alpha, \beta) = \alpha \beta - R^2$ we have:
\begin{equation}
\label{solution8}
V^{\pm} = \frac{2}{3}\lambda r \coth (2\lambda t)
\pm \frac{\sqrt{
4 \lambda^2 r^4 + 18 R^2 r \sinh^2(2\lambda t)
}}{3 r \sinh(2 \lambda t)}
\end{equation}
The expression for the density of energy is too large to print it here.
In the case $\lambda = 0$ we obtain:
\begin{equation}
\label{VG}
V^{\pm} = \frac{r}{3 t} \left(1 \pm \sqrt{1 + \frac{9 R^2 t}{r^3} } \right),
\end{equation}
\begin{equation}
\label{solution8Gtt}
G^{\pm}_{t t} = \frac{2}{3 t^2} \left( 
1 \pm \frac{1+\frac{9 R^2 t}{2 r^3} }{\sqrt{1 + \frac{9 R^2 t}{ r^3}}}
\right).
\end{equation}
In the case $R=0$ and positive sign we obtain the Friedmann solution with flat spatial distribution.
Due to the $t t$-equation of General Relativity~(\ref{ttEq}) and the expression~(\ref{solution8Gtt})
the density $\rho(t, r)$ is not a positive definite,
so the solution~(\ref{solution8}) does not always (everywhere) belong to General Relativity.

\paragraph{Solution 9}
In the case $F(\alpha, \beta) = \alpha^2 + \frac{A}{\lambda}  \beta$ we have:
\begin{equation}
\label{solution9}
V^{\pm} = \frac{\lambda r \sinh(2 \lambda t)}{3 \cosh^2(\lambda t)}
- \frac{A \tanh(\lambda t)}{2 \lambda \sqrt{r} \cosh(\lambda t)}
\pm \frac{\sqrt{
8 A r^{5/2} \cosh(\lambda t)
+ 3 A^2 r \lambda^{-2} \sinh^2(\lambda t)
}}{2\sqrt{3} r \cosh^2(\lambda t)}
\end{equation}
The expression for the density of energy is too large to print it here.
In the case $\lambda = 0$ we obtain:
\begin{equation}
\label{collapse1}
V^{\pm} = - \frac{A t}{2 \sqrt{r}} 
\pm 
\frac{\sqrt{8 A r^{5/2} + 3 A^2 r t^2}}{2 \sqrt{3} r},
\end{equation}
\begin{equation}
\label{solution9Gtt}
G^{\pm}_{t t} = \frac{A}{r^{3/2}} \mp \frac{A^2 t}{r \sqrt{ (8/3) A r^{5/2} + A^2 r t^2 }}.
\end{equation}
In the case of negative sign this solution is corresponding to collapsing Universe
with black hole which gravitational radius is growth with time (see the next section for details).
Due to the $t t$-equation of General Relativity~(\ref{ttEq}) and the expression~(\ref{solution9Gtt})
the density $\rho(t, r)$ is not a positive definite,
so the solution~(\ref{solution9}) does not always (everywhere) belong to General Relativity.
The negative density problem is considered further in the section ''Gravitational field equations''.
 
%*************************************************************
\subsection{The gravitational radius of a dust cloud and the radius of visible horizon}
\noindent
For the metric~(\ref{metric1}) the gravitational radius $r_g(t)$ of a dust cloud
and the radius of visible horizon are depends from the time $t$
and satisfy the following equation
\begin{equation}
\label{GRR}
1 - V(t, r_g(t))^2 = 0.
\end{equation}
For the solution~(\ref{VB}) with positive sign the equation~(\ref{GRR})
has two real non-negative roots $r^{(1)}_g(t)$ and $r^{(2)}_g(t)$.
This roots has the following asymptotic behaviour at $t \to \infty$:
\begin{equation}
r^{(1)}_g(t) \approx
\frac{R^3}{t^2} + \frac{4 R^6}{3 t^5} + \frac{28 R^9}{9 t^8}
+ O(t^{-11})
\end{equation}
\begin{equation}
r^{(2)}_g(t) \approx \frac{3 t}{2} 
- \sqrt{ \frac{3 R^3}{2 t} }
- \frac{R^3}{2 t^2}
- \sqrt{ \frac{25 R^9}{96 t^7} }
+ O(t^{-5})
\end{equation}
The root $r^{(1)}_g(t)$ is a gravitational radius of the dust cloud, it asymptotically decreases as $t^{-2}$.
The root $r^{(2)}_g(t)$ is a visible horizon, it asymptotically increases like in Friedmann solution.

For the solution~(\ref{VG}) with positive sing the equation~(\ref{GRR}) has two roots:
\begin{equation}
r^{(1)}_g(t) = \frac{3 t}{4} \left( 1 - \sqrt{1 - \frac{8 R^2}{3 t^2} } \right)
\approx
\frac{R^2}{t} +\frac{2 R^4}{3 t^3} + O(t^{-5});
\end{equation}
\begin{equation}
r^{(2)}_g(t) = \frac{3 t}{4} \left( 1 + \sqrt{1 - \frac{8 R^2}{3 t^2} } \right)
\approx
\frac{3 t}{2} - \frac{R^2}{t} - \frac{2 R^4}{3 t^3} + O(t^{-5}).
\end{equation}
The gravitational radius of different clouds have different asymptotics.
Solutions~(\ref{VB}) and~(\ref{VG}) has the same Friedmann asymptotic behavior for $r^{(2)}_g(t)$, but it has the different asymptotic behavior for $r^{(1)}_g(t)$: $R^3 / t^2$ vs $R^2 / t$.
The dust cloud~(\ref{VB}) is scattered faster than the dust cloud~(\ref{VG}).

Let us consider the asymptotic behaviour at $t \to +\infty$ of the solution~(\ref{collapse1}) with negative sign and $A \ne 0$:
\begin{equation}
V \approx - \frac{A t}{\sqrt{r}} - \frac{2 r}{3 t} + O(t^{-3}).
\end{equation}
The first term is responsible for the massive core of the collapsing dust cloud.
We rename it as follows
\begin{equation}
- \frac{A t}{\sqrt{r}} = - \sqrt{ \frac{2 \kappa M(t)}{r} },
\quad
M(t) = \frac{A^2 t^2}{2 \kappa}.
\end{equation}
The gravitational radius of this collapsing dust cloud asymptotically grows with time as $t^2$.
The negative linear by $r$ term is responsible for collapse of the Universe.

%*************************************************************
\subsection{A method of generating an infinite number of Tolman's like solutions}
\noindent
The generalization of the metric~(\ref{metric1}) to the case of a curved spatial section has the form:
\begin{equation}
\label{metric1f}
ds^2 = dt^2 - \left( \frac{ dr - V(t,r) \, dt }{W(t,r)} \right)^2
- r^2 \left( d\theta^2 + \sin^2 (\theta) \, d \varphi^2 \right).
\end{equation}
For the metric~(\ref{metric1f}), the system of equations~(\ref{eq1}) takes the following form
\begin{equation}
\label{metric1feq}
\frac{\partial V}{\partial t} 
+ \frac{1}{2 r} \frac{\partial}{\partial r} \left( r V^2 - \frac{4}{9} \lambda^2 r^3 \right) 
- \frac{W^2-1}{2 r} = 0,
\qquad
\frac{\partial W}{\partial t} + V \frac{\partial W}{\partial r} = 0.
\end{equation}
For the left side of the $t t$-equation of General Relativity we have the following expression:
\begin{equation}
G_{t t} - \frac{4}{3}\lambda^2 g_{t t}
= - \frac{2}{r} \left( \frac{\partial V}{\partial t}  + W \frac{\partial W }{\partial r} \right).
\end{equation}
Let us consider the following vacuum static solution of~(\ref{metric1feq}):
\begin{equation}
V(r) = \pm \sqrt{\frac{2 \kappa M}{r} + \frac{4}{9} \lambda^2 r^2 + W^2-1}, 
\quad
M = const,
\quad
W = const, 
\quad
\rho = 0.
\end{equation}
We pass to the synchronous coordinates $(t, r) \to (t, \xi)$:
\begin{equation}
\label{sc1}
\frac{\partial r}{\partial t} = V
\quad \to \quad
dr - V dt = \frac{\partial r}{\partial \xi} d \xi.
\end{equation}
\begin{equation}
ds^2 = dt^2 - \left( \frac{ \frac{\partial r}{\partial \xi} d \xi }{W} \right)^2
- r^2(t, \xi) \left( d\theta^2 + \sin^2 (\theta) \, d \varphi^2 \right).
\end{equation}
The dependence of $r$ on $\xi $ is arbitrary, only $(\partial r / \partial \xi) \ne 0$ is required.
If now we take arbitrary functions $M(\xi)$ and $W(\xi)$ instead of constant $M$ and $W$,
we get Tolman's solution~\cite{Tolman1934} with nonzero energy density:
\begin{equation}
G_{t t} - \frac{4}{3}\lambda^2 g_{t t}
= \frac{2 \kappa \frac{\partial M}{\partial \xi} }{r^2 \frac{\partial r}{\partial \xi} }.
\end{equation}
A similar program of actions can be performed with any solution 
$V_{[F]}(t, r)$ of the system~(\ref{metric1feq}).
This is the method of generating new Tolman type solutions.
For the second example of this method, let us consider the solution~(\ref{VB}), 
we pass to the synchronous coordinates $(t, r) \to (t, \xi)$ according to~(\ref{sc1}):
\begin{equation}
\frac{\partial r}{\partial t}^{\pm} = \frac{2 r}{3 t} \pm \frac{R}{t} \sqrt{ \frac{R}{r} }.
\end{equation}
If now we take arbitrary function $R(\xi)$ instead of constant $R$,
we obtain new Tolman type solution with the following energy density:
\begin{equation}
G_{t t}
=\frac{4}{3 t^2}
\left( 1 \pm \frac{3 R}{2 r} \sqrt{ \frac{R}{r} } \right)
\left( 1 \pm \frac{3 R \frac{\partial R}{\partial \xi}}{2 r \frac{\partial r}{\partial \xi} \sqrt{ \frac{R}{r} }  } \right)
\end{equation} 
As a third example, we take the solution~(\ref{VG}) and
pass to the synchronous coordinates $(t, r) \to (t, \xi)$ according to~(\ref{sc1}):
\begin{equation}
\frac{\partial r}{\partial t}^{\pm} = \frac{r}{3 t} \left(1 \pm \sqrt{1 + \frac{9 R^2 t}{r^3} } \right).
\end{equation}
If now we take arbitrary function $R(\xi)$ instead of constant $R$,
we obtain new Tolman type solution with the following density of energy:
\begin{equation}
G_{t t}
=
\frac{1}{3 t^2}
\left( 1 \pm \frac{1}{\sqrt{1 + \frac{9 t R^2}{r^3} }} \right)
\left( 1 \pm \sqrt{1 + \frac{9 t R^2}{r^3} }
+ \frac{6 t R \frac{\partial R}{\partial \xi} }{r^2 \frac{\partial r}{\partial \xi} }
\right).
\end{equation}
The fourth and fifth examples of new Tolman type solutions are given below (see~(\ref{t4}) and~(\ref{t5}) expressions).
Solutions of the system~(\ref{metric1feq}) are generators (generating functions) of new {\it Tolman type} solutions.

%*************************************************************
\section{Dust cloud in an expanding Universe filled with radiation}

\noindent
In this section we consider the problem of a spherically symmetric dust cloud
in the Universe filled with ultrarelativistic gas (radiation).
To take into account the gas pressure $p(t, r)$, we substitute to the right-hand side of the system~(\ref{eq1}) the following stress-energy tensor
\begin{equation}
\label{TP}
T_{\mu \nu} = \rho \, u_{\mu} u_{\nu} + \left( \varepsilon + p \right)  v_{\mu} v_{\nu} - p \, g_{\mu \nu}.
\end{equation}
Here $\rho$, $u_{\mu}$  is the density and the four-velocity of the ideal dust,
$\varepsilon$, $p$, $v_{\mu}$ is the density, pressure, and four-velocity of the ultrarelativistic gas (radiation).

First, we consider the case when the Universe filled only radiation at $\lambda = 0$.
The system~(\ref{eq1}) for the metric~(\ref{metric1}) take the form:
\begin{equation}
\label{eqVR}
\frac{\partial V}{\partial t} + \frac{1}{2 r} \frac{\partial}{\partial r} \left( r V^2 \right)
= - 4 \pi \kappa \, r \, p.
\end{equation}
\begin{equation}
v_{\mu}dx^{\mu} = dt, \quad
\frac{\partial p}{\partial r} = 0, \quad
\varepsilon = 3 p, \quad
\rho = 0.
\end{equation}
The solution is given by:
\begin{equation}
\label{VPR0}
V(t, r) = \frac{r}{2 t}, \quad
p(t) = \frac{1}{32 \pi \kappa t^2}, \quad
\varepsilon = 3 p, \quad
\rho = 0.
\end{equation}
According to~(\ref{fridman}) this corresponds to the Friedmann solution with scale factor $a(t) = A \sqrt{t}$.
Now we find particular solutions of the~(\ref{eqVR}) with the same radiation pressure $p(t)$,
but with a nonzero dust density $\rho$.
The simplest particular solution satisfying this criterion is
\begin{equation}
\label{VPR1}
V(t, r) = \frac{r}{2 t} \pm \sqrt{\frac{ 2k M(t)}{r}}, 
\quad
M(t) = \frac{Q^2}{2 \kappa \, t^{3/2}}, 
\quad
\rho = \frac{3 Q}{16\pi \kappa \, t^{7/4} r^{3/2}}.
\end{equation}
The gravitational radius of this dust cloud decreases asymptotically as $t^{-3/2}$.
The next particular solution is
\begin{equation}
\label{VPR2}
V(t, r) = \frac{r}{3 t} \left( 1 + \frac{1}{2} \sqrt{1 + \frac{R^2 t}{r^3}  } \right), 
\quad
\rho = \frac{
\frac{R^2 t}{r^3} + 2 \left( 1 - \sqrt{1 + \frac{R^2 t}{r^3}} \right)
}{
48 \pi \kappa \, t^2 \sqrt{1 + \frac{R^2 t}{r^3}}
}.
\end{equation}
The gravitational radius of this dust cloud decreases asymptotically as $t^{-1}$.
Thus, the gravitational radius of different clouds in the same expanding Universe
filled with radiation can have different asymptotics behavior.

\noindent
The particular solution~(\ref{VPR1}) generates the following family of Tolman's like solutions.
We pass to the synchronous coordinates $(t, r) \to (t, \xi)$:
\begin{equation}
\label{sc}
\frac{\partial r}{\partial t} = \frac{r(t,\xi)}{2 t} + \frac{Q}{t^{3/4}\sqrt{r(t,\xi)}},
\quad
dr - \frac{\partial r}{\partial t} dt = \frac{\partial r}{\partial \xi} d \xi.
\end{equation}
\begin{equation}
ds^2 = dt^2 - \left( \frac{\partial r}{\partial \xi}  \, d \xi \right)^2
- r^2(t, \xi) \left( d\theta^2 + \sin^2 (\theta) \, d \varphi^2 \right).
\end{equation}
If now we replace the constant $Q$ by an arbitrary function $Q (\xi)$, then we obtain a new family of Tolman's like solutions with the following dust density:
\begin{equation}
\label{t4}
\rho
= \frac{
2 r^{3/2} \frac{\partial Q}{\partial \xi} + Q 
\left( 4 t^{1/4} \frac{\partial Q}{\partial \xi} + 3 \sqrt{r} \frac{\partial r}{\partial \xi} \right)
}{
16 \pi \kappa \, t^{7/4} r^2 \frac{\partial r}{\partial \xi}
}.
\end{equation}
The particular solution~(\ref{VPR2}) generates the following family of Tolman's like solutions.
We pass to the synchronous coordinates $(t, r) \to (t, \xi)$:
\begin{equation}
\frac{\partial r}{\partial t} = 
\frac{r(t,\xi)}{3 t} \left( 1 + \frac{1}{2} \sqrt{1 + \frac{R^2 t}{r^3(t,\xi)}  } \right),
\quad
dr - \frac{\partial r}{\partial t} dt = \frac{\partial r}{\partial \xi} d \xi.
\end{equation}
If now we replace the constant $R$ by an arbitrary function $R (\xi)$, then we obtain a new family of Tolman's like solutions with the following dust density:
\begin{equation}
\label{t5}
\rho = \frac{
 r t R \frac{\partial R}{\partial \xi} \left( 2 + \sqrt{1 + \frac{t R^2}{r^3} } \right)
+ 3 r^3 \frac{\partial r}{\partial \xi} \left( \frac{t R^2}{r^3} - 2 \left( \sqrt{1+\frac{t R^2}{r^3} } - 1\right)\right)
}{
144 \pi \kappa\, t^2 r^3 \frac{\partial r}{\partial \xi} \sqrt{1+\frac{t R^2}{r^3} }.
}.
\end{equation} 

\noindent
The pressure~(\ref{VPR0}) corresponds to infinitely many particular solutions $V_{[F]} (t,r)$
which in implicit form are given by as a solution of equation:
\begin{equation}
\label{solution2}
F\left(   
\frac{\sqrt{r}}{t^{1/4}} \left( 2 t \, V_{[F]} (t,r) - r \right),
\frac{\sqrt{r}}{t^{3/4}} \left( 6 t \, V_{[F]} (t,r) - r \right)
 \right) = 0,
\end{equation}
here $F(\alpha, \beta)$ -- differentiable arbitrary function of two arguments.
The homogeneous solution $V = r / (2 t)$ is corresponding to the pure radiation case $\varepsilon = 3 p$.
The homogeneous solution $V = r / (6 t)$ is corresponding to the case $\varepsilon = (1/3) p$.
Since they both give exactly the same formula~(\ref{VPR0}) for pressure, the general solution contains an arbitrary combination of them.
If we are only interested in radiation, then we must choose only solutions from~(\ref{solution2}) that tend asymptotically to $V = r / (2 t)$. The negative density problem is considered further in the section ''Gravitational field equations''.

%*************************************************************
\section{Dust cloud in an expanding Universe filled with radiation and nonrelativistic gas}

\noindent
Let $a(t)$ is Friedmann scale factor from~(\ref{metricFLRW}),
according to~(\ref{radiusFLRW}) it corresponds to a radial velocity field $V(t, r) = r \, \dot{a}(t) / a(t)$.
We search a solution of the system~(\ref{eq1}) in the following form:
\begin{equation}
\label{Vrb}
V(t, r) = r \, \frac{\dot{a}(t) }{ a(t) } 
\pm \sqrt{ \frac{2 \kappa M(t)}{ r} + \frac{4}{9} \lambda^2 r^2 L(t)},
\end{equation}
\begin{equation}
\label{Prb}
p(t) = \frac{\Omega_{r}}{a^4(t)} + \frac{\Omega_{b}}{a^3(t)}.
\end{equation}
\begin{equation}
u_{\mu} dx^{\mu} = dt, \quad v_{\mu} dx^{\mu} = dt.
\end{equation}
In~(\ref{Prb}) the term with $\Omega_{r}$ is responsible for the radiation pressure, 
and the term with $\Omega_{b}$ is responsible for the pressure of the nonrelativistic (baryon) gas.
The system~(\ref{eq1}) reduces to the following ordinary differential equations:
\begin{equation}
\ddot{a}(t) = - \frac{\dot{a}^2(t)}{2 a(t)} 
- 4 \pi \kappa \left( \frac{\Omega_{b}}{a^2(t)} + \frac{ \Omega_{r}}{a^3(t)} \right)
+ \frac{2}{3} \lambda^2 a(t) (1 - L(t)),
\end{equation}
\begin{equation}
\dot{L}(t) = - 6 L(t) \frac{\dot{a}(t) }{ a(t) },
\end{equation}
\begin{equation}
\label{Mt}
\dot{M}(t) = - 3 M(t) \frac{\dot{a}(t) }{ a(t) }.
\end{equation}
Total density of dust, gas and radiation:
\begin{equation}
\rho + \varepsilon = 
\frac{3}{8 \pi \kappa} \left(
\frac{4 \lambda^2}{9} \left( L(t) - 1 \right)
+ \frac{\dot{a}^2(t)}{a^2(t)}
\pm \frac{\dot{a}(t)}{r a(t)}
\frac{
\frac{2 \kappa M(t)}{r} + \frac{8}{9} r^2 \lambda^2 L(t)
}{
\sqrt{ \frac{2 \kappa M(t)}{r} + \frac{4}{9} r^2 \lambda^2 L(t) }}
\right)
\end{equation}
It follows from~(\ref{Mt}) that the relative loss of the mass of the dust cloud~(\ref{Vrb}) is equal to the three Hubble constants $h(t)$:
\begin{equation}
h(t) = \frac{\dot{a}(t) }{ a(t) }, \quad
\frac{\dot{M}(t) }{ M(t) } = - 3 h(t).
\end{equation}
We note that a dust cloud of the form~(\ref{Vrb}) is only the simplest particular solution, there are other solutions in which the mass loss occurs according to a different law.

%*************************************************************
\section{Ultrarelativistic celestial body explosion}
\noindent
We consider the external problem of a spherically symmetric ultrarelativistic celestial body explosion (supernova, quasar).
Outside the celestial body, the space is filled with dust and an ultrarelativistic flux of explosion products (radiation).
They stress–energy tensor:
\begin{equation}
\label{TSE}
T_{\mu \nu} = \rho \, u_{\mu} u_{\nu} + N \, k_{\mu} k_{\nu}, \quad
g^{\mu \nu} T_{\mu \nu} = \rho, \quad
\nabla_{\mu} T^{\mu \nu} = 0.
\end{equation}
Here
$\rho$, $u^{\mu}$ is density and four-velocity of the dust, and
$N$, $k^{\mu}$  is density and four-velocity of the ultrarelativistic flux of explosion products.
The continuity and geodesic equations:
\begin{equation}
\nabla_{\mu} \left( \rho \, u^{\mu}\right) = 0, \quad
\nabla_{\mu} \left( N \, k^{\mu}\right) = 0, \quad
u^{\mu} \nabla_{\mu} u^{\nu} = 0, \quad
k^{\mu} \nabla_{\mu} k^{\nu} = 0.
\end{equation}
The $u^{\mu}$ is a timelike, and $k^{\mu}$ is an isotropic four-vectors:
\begin{equation}
\label{KK}
g_{\mu \nu} u^{\mu} u^{\nu} = 1, \quad
g_{\mu \nu} k^{\mu} k^{\nu} = 0.
\end{equation}
We use the following metric:
\begin{equation}
\label{metric1w}
ds^2 = dt^2 - \left( \frac{ dr - V(t,r) \, dt }{W(t,r)} \right)^2
- r^2 \left( d\theta^2 + \sin^2 (\theta) \, d \varphi^2 \right),
\end{equation}
and we use the following solution of equations~(\ref{KK}):
\begin{equation}
u^{\mu} \frac{\partial}{\partial x^{\mu}} = \frac{\partial}{\partial t} + V \frac{\partial}{\partial r},
\qquad
k^{\mu} \frac{\partial}{\partial x^{\mu}} = K \frac{\partial}{\partial t}
+ \left( V + W \right) K \frac{\partial}{\partial r}.
\end{equation}
The system of equations of General Relativity for the metric~(\ref{metric1w})
with the stress-energy tensor~(\ref{TSE})
takes the form:
\begin{equation}
\label{SE1}
\frac{\partial V}{\partial t}
 + \frac{1}{2r} \frac{\partial}{\partial r} \left( r V^2 - \frac{4}{9} r^3 \lambda^2 \right)
 - \frac{W^2-1}{2 r}
 =
 - 4\pi \kappa \, r \, N \, K^2,
\end{equation}
\begin{equation}
\label{SE2}
\frac{\partial W}{\partial t} + V \frac{\partial W}{\partial r}
  =
  4 \pi \kappa \, r \, N \, K^2,
\end{equation}
\begin{equation}
\label{SE3}
\frac{\partial K}{\partial t}
 + K \frac{\partial V}{\partial r}
 + \left( V + W \right) \frac{\partial K}{\partial r}
 =
 4 \pi \kappa \frac{r \, N \, K^3}{W},
\end{equation}
\begin{equation}
\label{SE4}
\frac{\partial N}{\partial t} + \frac{V + W}{r^2} \frac{\partial}{\partial r} \left( r^2 \, N \right) = 0.
\end{equation}
And from $t t$-equation of General Relativity for the density $\rho$ we have:
\begin{equation}
\rho = \frac{1}{8\pi \kappa \, r^2} \frac{\partial}{\partial r}
\left( r \left( V^2 - W^2 + 1 - \frac{4}{9} r^2 \lambda^2 \right) \right)
- \left(1 + \frac{V}{W} \right) N \, K^2.
\end{equation}
The system of equations~(\ref{SE1}, \ref{SE2}, \ref{SE3}, \ref{SE4}) describes
a spherically symmetric gravitational field outside the celestial body
considering the ultrarelativistic radiation flux coming from it.

%*************************************************************
\section{Space-Time structure}
\noindent
A metric of the form~(\ref{metric1}) is not only the key to the unification
of various solutions (Schwarzschild, de Sitter, Friedman, etc.), but also allows,
for example, to write the equations of the hydrodynamic problem most simply,
and then trace the fate of the falling matter under the horizon of black hole~\cite{Ruban2014}.
A metric of the form~(\ref{metric1}) is necessary for the realization of the
correct limit passage from General Relativity to the problems of classical physics.
For example, in classical mechanics, the Lagrangian of a free particle moving in
a velocity field $V^i$ has the following form:
\begin{equation}
L = \frac{1}{2} m \gamma_{i j} 
\left( \frac{dx}{dt}^i - V^i \right) \left( \frac{dx}{dt}^j - V^j \right).
\end{equation}
It exactly corresponds to the non-relativistic motion in the gravitational field of the form~(\ref{metric1}) if it is written in a slightly more general form~\cite{Burlankov2006}:
\begin{equation}
\label{metric2}
ds^2 = dt^2 - \gamma_{i j} \left( dx^i - V^i dt \right) \left( dx^j - V^j dt \right).
\end{equation}
The Burlankov metric~(\ref{metric2}) is a special case of the Arnovitt-Deser-Misner metric~\cite{ADM1959}.
The local space-time structure of the four-dimensional space of events is given by tetrad
which contains three covector fields (triad):
$e^{(1)} = e^{(1)}_{\mu} dx^{\mu}$,
$e^{(2)} = e^{(2)}_{\mu} dx^{\mu}$,
$e^{(3)} = e^{(3)}_{\mu} dx^{\mu}$
which defining a basis in the cotangent bundle of the three-dimensional spatial distribution,
and one covector field $e^{(0)} = e^{(0)}_{\mu} dx^{\mu}$ defining the differential form
of time~\cite{Sardanashvili2011}.
The time is transversal to the spatial distribution.
Taken together, $e^{(0)}$, $e^{(1)}$, $e^{(2)}$, $e^{(3)}$ give a basis in the linear reference frame bundle
with the Lorentz group, and the metric:
\begin{equation}
g_{\mu \nu} = e^{(0)}_{\mu} e^{(0)}_{\nu} - e^{(1)}_{\mu} e^{(1)}_{\nu}
 - e^{(2)}_{\mu} e^{(2)}_{\nu} - e^{(3)}_{\mu} e^{(3)}_{\nu}.
\end{equation}
\begin{equation}
\label{norma}
g^{\mu \nu} e^{(0)}_{\mu} e^{(0)}_{\nu} = +1, \quad
g^{\mu \nu} e^{(1)}_{\mu} e^{(1)}_{\nu} = -1, \quad
g^{\mu \nu} e^{(2)}_{\mu} e^{(2)}_{\nu} = -1, \quad
g^{\mu \nu} e^{(3)}_{\mu} e^{(3)}_{\nu} = -1.
\end{equation}
The Burlankov's metric~(\ref{metric2}) is corresponding to the case of exact differential form of the time:
\begin{equation}
\label{e0}
d e^{(0)} = 0, \qquad
\frac{\partial e^{(0)}_{\nu}}{\partial x^{\mu}} - \frac{\partial e^{(0)}_{\mu}}{\partial x^{\nu}} = 0, \qquad
e^{(0)} = \frac{\partial t}{\partial x^{\mu}} dx^{\mu}, \qquad
\oint e^{(0)} = 0.
\end{equation}
In this case, the integral manifolds of the local three-dimensional spatial distribution
form a foliation of the event space, whose sheets are the hypersurfaces $t(x) = const$.
Physically, the condition~(\ref{e0}) makes possible the integral (nonlocal)
time synchronization: the integral of $e^{(0)}$
depends only on the initial and final points, but not on the path of integration.
From~(\ref{norma}) and~(\ref{e0}) the Hamilton-Jacobi equation is obtained
for the time $t(x)$ used in the Burlankov metric:
\begin{equation}
\label{HY}
g^{\mu \nu} \frac{\partial t}{\partial x^{\mu}} \frac{\partial t}{\partial x^{\nu}} = 1.
\end{equation}
In the limit passage from General Relativity to nonrelativistic physics,
the function $t(x)$ satisfying the Hamilton-Jacobi equation
goes over to Newton's time in classical mechanics,
time in the Schrödinger equation in quantum mechanics,
and time in the Euler equation in hydrodynamics.
That's why the hydrodynamic problems~\cite{Ruban2014} in the Painlevé--Gullstrand coordinates are much easier to formulate.

%*************************************************************
\section{Gravitational field equations}
\noindent
Let us consider the problem of negative energy density arising in the General Relativity.
According to the solution~(\ref{generalSolution}) and solution~(\ref{solution2}) of any arbitrarily given function $F(\alpha, \beta)$ there corresponds a particular solution $V_{[F]} (t, r)$.
We denote $V_{[F^{+}]} (t, r)$,  $V_{[F^{0}]} (t, r)$, $V_{[F^{-}]} (t, r)$ solutions,
which make the left part of the $t t$-equation of General Relativity respectively positive, zero or negative.
On the right-hand side of the $t t$-equation of General Relativity there is a positively determined
energy density of regular matter, so the solutions $V_{[F^{-}]} (t, r)$ in General Relativity are forbidden at all.
However, from the geometric point of view, there is no fundamental difference between solutions with a positive or negative component $G_{t t}$ of the Einstein-Hilbert tensor.
The restriction of $G_{tt} \ge 0$ looks artificial from the geometric point of view.
It is interesting to change the equations of the gravitational field so that any sign $G_{t t}$ is allowed.
We find an extremum of Hilbert action not on arbitrary world manifolds as it is done in General Relativity,
but only on those manifolds that admit the exact differential form of time.
According to~(\ref{e0}) the following representation is admissible for the metric of such manifolds
\begin{equation}
\label{metric3}
g_{\mu \nu} = \frac{\partial t}{\partial x^{\mu}} \frac{\partial t}{\partial x^{\nu}} - e^{(1)}_{\mu} e^{(1)}_{\nu}
 - e^{(2)}_{\mu} e^{(2)}_{\nu} - e^{(3)}_{\mu} e^{(3)}_{\nu}.
\end{equation}
Keeping exact the differential form of time, we vary the Hilbert action in thirteen fields:
$t(x)$, $e^{(1)}_{\mu}(x)$, $e^{(2)}_{\mu}(x)$, $e^{(3)}_{\mu}(x)$:
\begin{equation}
\delta S = \frac{1}{2} \int \left( T_{\mu \nu} + \frac{\lambda^2}{6\pi \kappa} g_{\mu \nu} - \frac{1}{8 \pi \kappa} G_{\mu \nu}  \right) 
\delta g^{\mu \nu} \sqrt{-g} \, d_4 x.
\end{equation}
We obtain the following system of equations of the gravitational field:
\begin{equation}
\label{eq12}
\left( T^{\mu \nu} + \frac{\lambda^2}{6\pi \kappa} g^{\mu \nu} - \frac{1}{8 \pi \kappa} G^{\mu \nu}  \right) e^{(i)}_{\nu} = 0,
\end{equation}
\begin{equation}
\label{eq13}
\nabla_{\mu} \left( \left( T^{\mu \nu} + \frac{\lambda^2}{6\pi \kappa} g^{\mu \nu} - \frac{1}{8 \pi \kappa} G^{\mu \nu} \right) 
\frac{\partial t}{\partial x^{\nu}} \right)  = 0.
\end{equation}
Due to local {\bf \it SO}(3) symmetry of the triad $e^{(1)}_{\mu}$, $e^{(2)}_{\mu}$, $e^{(3)}_{\mu}$ of the twelve equations~(\ref{eq12}) only nine are linearly independent.
The equation~(\ref{eq13}) is the energy conservation law.
Let $P^{\mu}$ be the conserved energy-momentum density current:
\begin{equation}
\label{P}
P^{\mu} = \left( T^{\mu \nu} + \frac{\lambda^2}{6\pi \kappa} g^{\mu \nu} - \frac{1}{8 \pi \kappa} G^{\mu \nu} \right) 
\frac{\partial t}{\partial x^{\nu}},
\qquad
\nabla_{\mu} P^{\mu} = 0.
\end{equation}
If the nine linearly independent equations~(\ref{eq12}) are satisfied, then the equation~(\ref{eq13}) is satisfied automatically due to the Hilbert identities. For the metric~(\ref{metric1}) and the stress-energy tensor~(\ref{TP}) for the nonzero components of $P^{\mu}$ we have:
\begin{equation}
P^{t} = \epsilon, \quad
P^{r} = V \epsilon, \quad
\epsilon = \rho + \varepsilon + \frac{\lambda^2}{6 \pi \kappa }
- \frac{1}{8 \pi \kappa \, r^2} \frac{\partial}{\partial r} \left( r V^2 \right),
\end{equation}
\begin{equation}
\frac{\partial \epsilon}{\partial t} + \frac{1}{r^2} \frac{\partial}{\partial r} \left( r^2 V \epsilon \right) = 0.
\end{equation}
The nine linearly independent equations~(\ref{eq12}) are covariant formulation of Burlankov's theory of gravity~\cite{Burlankov2006}.
The solutions $V_{[F^{-}]}(t, r)$ belong to the Burlankov's theory of gravity.
The $t t$-equation of the General Relativity from the point of view of Burlankov's gravitation theory
is a condition like $H = 0$, where $H$ is the Hamiltonian. 
Thus, the General Relativity is responsible for describing the zero modes only.

%*************************************************************
\section{Conclusion}
\noindent
The author is grateful to Viktor Konstantinovich Dubrovich and Dmitry Evgenievich Burlankov.

%*************************************************************
\section{Post Scriptum}
After this article was posted to the {\bf arXiv}, I was contacted by Michel Mizony and he informed me that he think I received results very similar to the results received by him.
Michel investigated the following metric:
\begin{equation}
ds^2 = dt^2 - \frac{(dr - v(t,r) dt)^2}{1 + v^2(t,r) - 2 \Phi(t,r)}
- r^2 \left( d \theta^2 + \sin^2 (\theta) d \varphi^2 \right).
\end{equation}
I am pleased to add references to his works~\cite{Mizony1, Mizony2, Mizony3}.

%****************************************************************

\end{document}